# A Novel Design for Quantum-dot Cellular Automata Cells and Full Adders

Mostafa Rahimi Azghadi[*], O. Kavehei, K. Navi
Department of Electrical and Computer Engineering,
Shahid Beheshti University, Tehran, Iran.
[*]m_rahimi@sbu.ac.ir and mostafa@eleceng.adelaide.edu.au

**Abstract-** Quantum dot Cellular Automata (QCA) is a novel and potentially attractive technology for implementing computing architectures at the nano-scale. The basic Boolean primitive in QCA is the majority gate. In this paper we present a novel design for QCA cells and another possible and unconventional scheme for majority gates. By applying these items, the hardware requirements for a QCA design can be reduced and circuits can be simpler in level and gate counts. As an example, a 1-bit QCA adder is constructed by applying our new scheme and is compared to the other existing implementation. Beside, some Boolean functions are expressed as examples and it has been shown, how our reduction method by using new proposed item, decreases gate counts and levels in comparison to the other previous methods.

**Key words:** QCA, three cube QCA cell, five-input majority gate, logical function, logical simplification, QCA Full Adder, hardware reduction.

## I. Introduction

Quantum Cellular Automata (QCA) is a nanotechnology that has recently been recognized as one of the top six emerging technologies with potential applications in future computers [1][2]. It has gained significant popularity in recent years. This is mainly due to rising interest in creating computing devices and implementing any logical function with that. The basic building block of QCA circuit is majority gate; hence, efficiently constructing QCA circuits using majority gates has attracted a lot of attentions [3]. Several studies have reported that QCA can be used to design general purpose computational and memory circuits [4]. QCA is expected to achieve high device density, very high clock frequency and extremely low power consumption [2].

In recent years the development of integrated circuits has been essentially based on scaling down that is, increasing the element density on the wafer. Scaling down of complementary metal oxide semiconductor CMOS circuits, however, has its limits. Above a certain element density various physical phenomena, including quantum effects, conspire to make transistor operation difficult if not impossible. If a new technology is to be created for devices of nanometer scale, new design principles are necessary. One promising approach is to move to a transistorless cellular architecture based on interacting quantum dots, quantum dot cellular automata QCA [5][14]**.**

At this time, it is unclear whether or not this technology will replace such a firmly embedded technology as CMOS, but investigations into modeling and design have demonstrated that QCA has many powerful features some of which are not available in CMOS [6][7]. Although many fabrication challenges have still to be overcome [7][2], the simple design nature of QCA makes it attractive for investigation of new circuit topologies [8]. QCA topologies are not simple translations of standard circuit layouts; new ideas for translating standard logic units into QCA are needed. One of the interesting features of QCA is that there is no fixed connection strategy, and hence, it should be possible to apply optimization algorithms such as genetic algorithms to minimize the number of cells in a design. In [9] a logic optimization method is introduced and hardware saving using genetic algorithm is performed.

In this paper we propose another possible design for majority gates resulting in a novel scheme for QCA cells, to facilitate simplifying logical functions. By applying this form of majority gate, we can simplify logical functions and achieve improved results. For example a 1-bit QCA adder is constructed only with three gates (two different forms of majority gates and only one inverter). In comparison to other existing implementation this method has demonstrated interesting results. Beside, some Boolean functions are expressed as examples and it has been shown, how our reduction method by applying new proposed item, decreases gate counts and levels. We will show and discuss that using of the proposed items can be efficient in designing majority gate based circuits. Furthermore, we will discuss the construction of primitive components for circuit designing in QCA technology such as QCA wire, majority gate and inverter using this novel form of quantum cell.

This paper is organized as follows. In section II, we provide a brief background to QCA technology. In section III, we develop a new design of QCA cells and illustrate another possible form of majority gates. Section IV, as an example, introduces a 1-bit QCA adder constructed with only two majority gates and one inverter and compares it to other exist implementations. In addition, in this section we express some Boolean function as examples and it has been shown, how our reduction method decrease gate counts and levels in comparison to the other previous methods. Finally, we conclude this paper in section V.

## II. Background Materials
### A. Quantum cells

The principle of quantum cellular automata was first proposed by Lent et al [16]. In order to implement a system that encodes information in the form of electron position it becomes necessary to construct a vessel in which an electron can be trapped and "counted" as there or not there. A quantum dot does just this by establishing a region of low potential surrounded by a ring of high potential.

In ordinary form, QCA technology is based on the interaction of bi-stable QCA cells constructed from four quantum dots. A schematic of a basic cell is shown in Fig. 1(a). The cell is charged with two free electrons, which are able to tunnel between adjacent dots. These electrons tend to occupy antipodal sites as a result of their mutual electrostatic repulsion. Thus, there exist two equivalent energetically minimal arrangements of the two electrons in the QCA cell, as shown in Fig. 1(b and c). These two arrangements are denoted as cell polarization. And by using cell polarization to represent logic "1" and to represent logic "0" binary information is encoded in the charge configuration of the QCA cell [10].

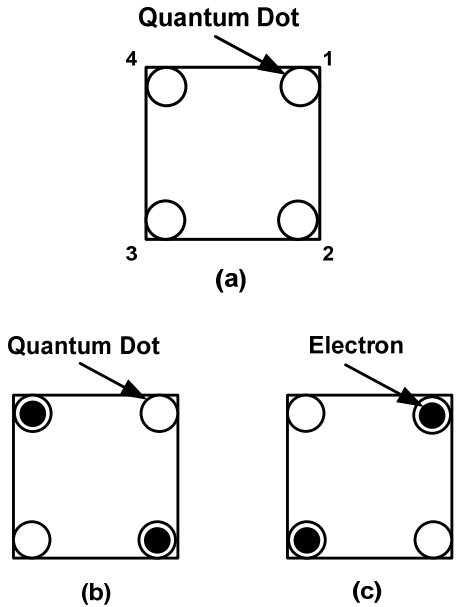

As noted above, Coulomb repulsion causes the electrons to occupy antipodal sites, the ground state charge distribution may have the electrons aligned along either of the two diagonal axes shown in Fig. 1 (b and c). In [6] the cell polarization has been defined, a quantity which measures the extent to which the charge distribution is aligned along one of these axes. The polarization is defined as Eq. 1.

$$P \equiv \frac{(\rho_2 + \rho_4) - (\rho_1 + \rho_3)}{\rho_1 + \rho_2 + \rho_3 + \rho_4} \quad (1)$$

Where $\rho_i$ denotes the electronic charge at dot i. With respect to this equation, electrons exactly localized on sites two and four will result in P=+1, while electrons on sites one and three yield P=-1. In next section we will use a similar expression for our approach presentation. In [6], the standard cell parameters have been discussed. And a simple model of the standard quantum cell has been employed, representing the quantum dots as sites and ignoring any degrees of freedom internal to the dot. Moreover, a second-quantized Hubbard-type Hamiltonian in the basis of two-particle site kets for this model has been used. Finally, a function for calculating cell-cell response has been expressed, because for using of cellular automata type architectures, the state of a cell must be strongly influenced by the states of neighboring cells.

### B. Nanoelectronic Circuits based on three-input Majority gates and Inverters

Any QCA circuit can be efficiently built using only majority gates and inverters. As shown in Fig. 2 (a), an ordinary QCA gate implementing the majority function is as follows.

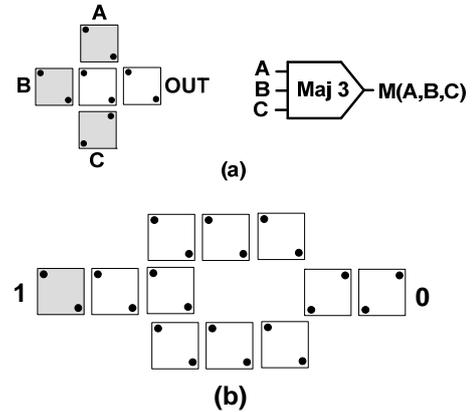

Fig.1. Basic QCA cell and binary encoding. (a) Schematic of the basic cell constructed from four quantum dots. Coulombic repulsion causes the electrons to occupy antipodal sites within the cell. These two bistable states result in cell polarizations of (b) P = +1 (Binary1), (c) P = -1 (Binary 0)

Fig.2. Primitive components. (a) A QCA majority gate. The states of the center and right cells are always the same as the state of the majority of the three input neighbors. (b) A QCA inverter. The signal comes in from the left, splits into two parallel wires, and is inverted at the point of convergence.

Assuming the inputs are A, B and C, the logic function of a majority gate is shown in Eq. 2.

M (A, B, C) =AB+BC+AC            (2)

As illustrated in Fig. 2 (a and b), each QCA majority gate in normal form requires only five QCA cells. And every QCA inverter gate can be implemented by 11 quantum cells. As it is shown in Eq. (3a) and (3b), for generating an AND gate or an OR gate with majority we can fix the polarization of one input to a constant logic '0' or logic '1'. Hence, QCA circuit is based on majority gate-based circuits instead of AND/OR/Inverter gate-based circuits [3][11].

M (A, B, 0) = AB           (3a)
M (A, B, 1) = A+B          (3b)

Other important component in QCA designing is wire. In a QCA wire, the binary signal propagates from input to output because of the electrostatic interactions between cells. Since the polarization of each cell tends to align with that of its neighbors, a linear arrangement of standard cells can be used to transmit binary information from one point to another. In this wire, all of free cells align in the same direction as the driving cell (input cell), so the information contained in the state of the input is transmitted down the wire. Beside, the distance between dots and between cells is a key parameter giving Coulombic effect in QCA application in conventional form [17]. A QCA wire is shown in Fig. 3. Moreover in a QCA wire All of the computational power is provided by the Coulomb interaction between cells, and there is no electrical current flow between cells and hence no power dissipation.

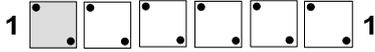

**Fig.3. QCA wire.** The binary signal propagates from input to output because of the electrostatic interactions between cells.

## III. Five-input Majority gate and novel design for QCA cell

A five pins majority gate must have five inputs and one output. In ordinary form of QCA cells, a cell can take effect from three around sides (bottom, up and left) and out the result function from another side (right). Hence this form of cell is proper for three-input majority implementation, but for a five-input majority gate we need a different design of cells, that can affect from five sides and transmit its polarization from another side. Hence we need to have a new structure in QCA cells including five inputs and one output. A new scheme for QCA cells is presented here. In the proposed design a QCA cell is a structure comprised of eight quantum-dots arranged in a cube pattern as shown in Fig. 4. We propose this structure for compatibility with five pins majority gate.

As mentioned, QCA uses the positions of electrons in quantum dots to represent binary values '0' and '1'.

In the proposed design, we have the same conditions. Fig. 5 illustrates a QCA cell with eight quantum dots. Four electrons occupy each cell. Each electron is free to tunnel between dots within one cell, but cannot leave the cell. The four electrons within each cell repel each other to diagonally opposite corners of the cell. This leaves only two stable states for each cell. These two states are used to represent logic values.

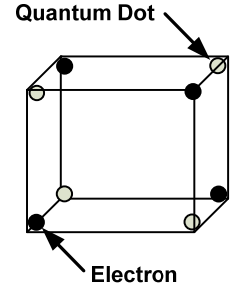

**Fig.4. Three cube QCA cell.** Schematic of a novel QCA cell comprised of eight quantum dots. Coulombic repulsion causes the electrons to occupy antipodal sites within the cell.

As shown in Fig. 5, in our design we determine every corner and thus every quantum dot with a number, which is between one up to eight. The occupation of dots in the corners with numbers one, three, six and eight represent logic '1', as shown in Fig. 5(a). In this case, the QCA cell is said to be polarized to +1. In Fig. 5(b), similarly, the occupation of two, four, five and seven dots represents logic '0'. In this case, the QCA cell is said to be polarized to -1.

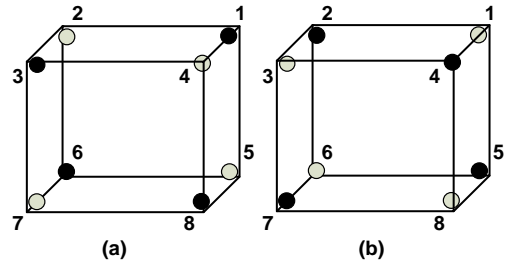

**Fig.5. Binary encoding.** The four electrons within each cell repel each other to diagonally opposite corners of the cell. This leaves only two stable states for each cell. (a) P = +1(Binary 1), (b) P = -1(Binary 0)

Like ordinary form of QCA cells, in proposed form of QCA cells, an equation for cell polarization can be defined as follows:

$$P \equiv \frac{(\rho_1 + \rho_3 + \rho_6 + \rho_8) - (\rho_2 + \rho_4 + \rho_5 + \rho_7)}{\rho_1 + \rho_2 + \rho_3 + \rho_4 + \rho_5 + \rho_6 + \rho_7 + \rho_8} \quad (4)$$

In Eq. 4, $\rho_i$ denotes the electronic charge at site i. The charge on each site can be calculated by finding the expectation value of the number operator in the ground state.

In this equation, electrons completely localized on sites one, three, six and eight polarization result in binary one (P=+1), and while electrons on sites two, four, five and seven yield binary zero (P=-1). Beside, polarization can get a value between these two complete polarization magnitudes.

This kind of QCA cells has a similar treatment to ordinary QCA cells. Hence, we can generalize all of the problems and questions around that, to new QCA cell form.

In this new form of QCA cell, moreover we can have primitive components for designing every function only with QCA cells. These components are QCA wire, majority and inverter gates. These components in our approach can be represented as explained in Section A to C.

**A. QCA wires**

With respect to new design of QCA cells, QCA wires can be represented as illustrated in Fig. 6. Adjacent QCA cells interact in an attempt to settle to a ground state determined by the current state of the inputs. Since the polarization of each cell tends to align with that of its neighbors. This is most clear in the case of the QCA wire shown in Fig. 6. The polarization of the input cell is propagated down the wire, as a result of the system attempting to settle to a ground state. Any cells along the wire that are anti-polarized to the input would be at a higher energy level, and would soon settle to the correct ground state.

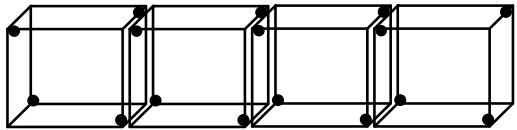

**Fig.6. QCA wire**. The binary signal propagates from input to output because of the electrostatic interactions between cells.

**B. QCA inverter**

Two standard cells in a diagonal orientation are geometrically similar to two rotated cells in a horizontal orientation. For this reason, standard cells in a diagonal orientation tend to align in opposite polarization directions as in the inverter chain [6]. Computation with QCA is accomplished by designing QCA layouts, which exhibit the desired interaction of states. Consider the arrangements in Fig. 7, demonstrating the QCA implementation of an inverter. Similarly to ordinary QCA inverter, in this form we have 11 QCA cells.

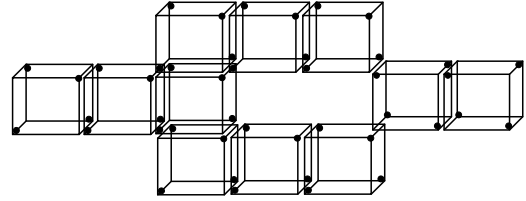

**Fig.7. QCA Inverter**. The signal comes in from the left, splits into two parallel wires, and is inverted at the point of convergence. An inverter gate can be implemented using 11 new proposed QCA cells.

**C. Five-input Majority gate**

Majority is a voter. In a new structure, a majority gate can be implemented as shown in Figure 8(a). Polarization of input cells are fixed and middle cell and out cell are free. Inputs from five around cells affect on the middle cell, and it determines polarization of the output cell. This structure makes majority decision. The majority voting logic function can be expressed in terms of fundamental Boolean operator as Eq. 5.

M(A,B,C,D,E)=ABC+ABD+ABE+ACD+ACE+ADE+
BCD+BCE+BDE+CDE                           (5)

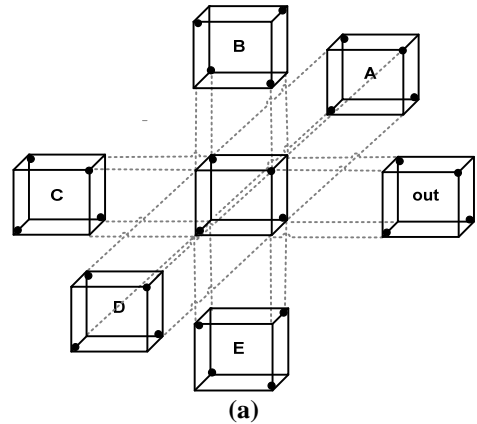

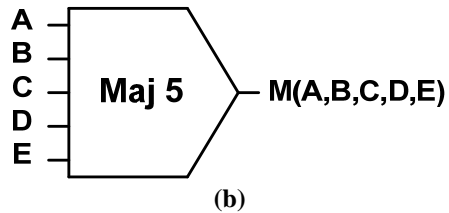

**Fig.8. Majority gate.** (a) The states of the center and out cells are always the same as the state of the majority of the five input neighbors. (b) Schematic symbol for the majority gate.

A schematic symbol of a five pins majority gate is shown in Figure 8(b). We can implement a three-input AND gate and also a three-input OR gate using this majority gate. These functions are shown in Eq. 6 (a and b).

M(A,B,C,0,0)= ABC        (6a)
M(A,B,C,1,1)= A+B+C      (6b)

### IV. One bit QCA full adder

Here, we will apply the proposed majority construction to design QCA full adder. First, we present other implementations and then our design is presented.
A 1-bit full adder is defined as follows:
Inputs: Operand bits (A, B) and carry bit is shown as C.
Outputs: Sum bit and carry out.

#### A. A QCA Adder with eight gates

In a classic design in [6], a full adder with five, three-input majority gates and three inverters has been implemented. This design is shown in Fig. 9. We can simplify this design and reduce one majority gate simply, for example we can replace $M(A,B,\overline{C})$ with only $\overline{C}$. Thus, an easier form for this QCA adder can be implemented using four majority gates and three inverters. It must be notified that in the following designs, a single bit full adder is implemented using only inverter and majority gates, whereas in order to implement the some function based on AND and OR gates, the number of gates involved will be increased.

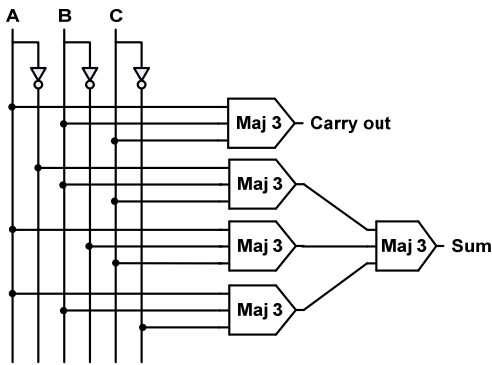

**Fig.9. One-bit QCA Adder with eight gates implemented in [6].**
Five, three-input majority gate and three inverter gates.

#### B. A QCA Adder with five gates

In [10] a method for majority logic reduction for QCA circuits has been presented and another form of QCA adder is expressed. This design is shown in Fig. 10. In this design we have three majority gates and two inverters. Beside, this design has been used in [15], recently.

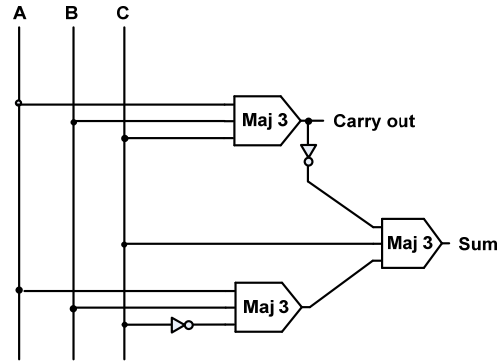

**Fig.10. One-bit QCA Adder with five gates implemented in [10].**
Three majority gates and two inverters.

In [10], a method for reducing the number of majority gates required for computing three variable Boolean functions is developed to facilitate the conversion of sum-of-products expression into QCA majority logic. Thirteen standard functions are introduced to represent all three variable Boolean functions and the simplified majority expressions corresponding to these standard functions are presented. In this paper a novel method for using these standard functions has been described to convert the sum-of-products expression to majority logic. By applying this method a considerable gate counts reduction will be achieved as illustrated in table I.

#### C. A QCA Adder only with three gates

In the proposed design, we simplify the QCA adder and reduce the number of both majority gates and inverters. As shown in Fig. 11, in our design the one bit full adder comprises only two majority gates and one inverter.

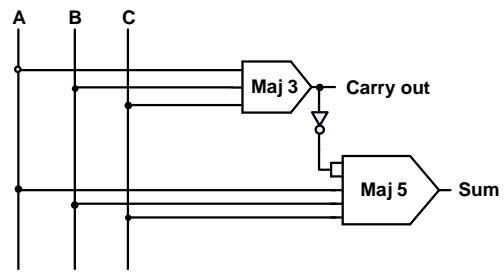

**Fig.11. One-bit QCA Adder with only three gates implemented using our proposed method.** One, five-input majority gate, one three-input majority gate and one inverter.

As noted above, proposed method represents a considerable reduction in gate counts compared to the other existing methods [6] [10], and retains the simple clocking scheme. This reduction is shown in table I.

Table I. comparison of QCA Adders

| QCA Adder | Complexity of one bit adder | Clocking |
|---|---|---|
| Our reduction method | Two majority gates and one Inverter | simple |
| Previous design [10] | Three majority gates and two inverters | simple |
| Previous design[6] | Five majority gates and three inverters | simple |

As is mentioned, the proposed item is very useful in circuit designing. In so many cases the standard Boolean function can be simplified and we are able to convert the Sum of Products expression to five- and three-input majority based design. In the following some examples are demonstrated, in which some three variables Boolean functions realized with Sum of Products are implemented using the five- and three-input majority gates. These examples are cited in [10] and [12], and we compare their approach to our's. These examples are shown in table II. In [12] a logic optimization method for majority gate based Nanoelectronic circuits using a novel mapping and sharing scheme is proposed to achieve simple synthesized circuits and high synthesis speed. In [12] they have tried to optimize the previous work presented in [3] [13], and the presented method has a considerable improvement in comparison to those works. It is obvious from table II the proposed item in this paper can be very efficient in majority and inverter gate counts reduction.

Table II. Majority expression of some three variable Boolean functions

| Minterms | Previous Approach [10][12] | Our design Approach |
|---|---|---|
| $\sum(7)$ | M(M(A,B,0),C,0) | M5(0,0,A,B,C) |
| $\sum(3,4,5,6,7)$ | M(M(B,C,0),A,1) | M5(A,A,B,C,1) |
| $\sum(3,6,7)$ | M(0,B,M(A,C,1)) | M5(A,B,B,C,0) |
| $\sum(1,2,3,4,5,6,7)$ | M(M(A,B,1),C,1) | M5(A,B,C,1,1) |
| $\sum(1,2,7)$ | M(M(A,B,C'),M(A,B',C),M(A',B,0)) | M5(M(A,B,C)',M5(A,A,B,C,1),A,B,C) |
| $\sum(0,3,5,6,7)$ | M(M(A,B,0),M(A',B',C),M(A,C',1)) | M(M5(A,B,B,C,C),1,M5(A,B,C,1,1)') |

We compared our approach to the method presented in [10][12] and in this comparison we achieve some considerable and efficient improvements. These important results illustrated in table III. The reduction is in the number of circuit levels, majority gate counts and inverter gate counts. Table III shows that our method gains advantages in designing boolean circuits. In the first function presented in table III, level counts reduced 50 percents. In addition, we replace two three-input majority gate with a five-input majority gate and hence gate counts is reduced. In the second function, the number of levels are not changed but inverter counts has a good reduction. Also, the number of majority gates has been reduced. In our approach we have used both five-input majority gates and three-input majority gates simultaneously. In the second function, in comparison to previous design number of inverters has a considerable reduction. And thus, gate counts have a good reduction.

Table III. Optimization results and comparison

| | $\sum(3,4,5,6,7)$ | | | |
|---|---|---|---|---|
| | Levels | Inverters | Majority gates | gates |
| **Previous works [10][12]** | 2 | 0 | 2 maj3 | 2 |
| **Proposed method** | 1 | 0 | 1 maj5 | 1 |
| | $\sum(0,3,5,6,7)$ | | | |
| | Levels | Inverters | Majority gates | gates |
| **Previous works [10][12]** | 2 | 3 | 4 maj3 | 7 |
| **Proposed method** | 2 | 1 | 1 maj3, 2 maj5 | 4 |

## V. Conclusion

In this paper, a novel scheme for QCA cell has been introduced to construct a five-input majority gate in order to make efficient QCA majority logic design. Furthermore, we have introduced a new strategy for simplifying boolean functions based on five-input majority gate. The study has had a direct application to the design of logic functions in QCA where the logic primitives are three- and five-input majority gate. As a case study, the proposed technique is used to develop a 1-bit QCA adder that is constructed with only two majority gates and one inverter. The QCA adder has been compared with another existing implementation of QCA adders. Also, some examples illustrates that our method has considerable influence on circuits simplifying as our proposed item reduces the level counts and the number of majority and inverter. It is expected that the new scheme for QCA cells and the new form of majority gate and the reduction method presented in this paper produces significant improvement in majority gate based nanoelectronic circuits and reduces chip area and increases speed for many future QCA architectures.